\documentclass[sigconf]{acmart}
\usepackage{listings}

\makeatletter
\def\@ACM@checkaffil{
    \if@ACM@instpresent\else
    \ClassWarningNoLine{\@classname}{No institution present for an affiliation}%
    \fi
    \if@ACM@citypresent\else
    \ClassWarningNoLine{\@classname}{No city present for an affiliation}%
    \fi
    \if@ACM@countrypresent\else
        \ClassWarningNoLine{\@classname}{No country present for an affiliation}%
    \fi
}
\makeatother

\begin{document}
\fancyhead{}
\def\thetitle{Just-in-Time autotuning}
\title{\thetitle}

\author{Elian Morel\dag, Camille Coti\ddag}
\affiliation{\small{
\dag École Polytechnique, Palaiseau, France
\and
\ddag École de Technologie Supérieure, Montréal, Canada
}}

\date{}

\begin{abstract}
Performance portability is a major concern on current architectures. One way to 
achieve it is by using autotuning. In this paper, we are presenting how we exten
ded a just-in-time compilation infrastructure to introduce autotuning capabiliti
es triggered at run-time. When a function is executed, the first iterations opti
mize it, and once the best solution has been found, it is used for subsequent calls to the function. This just-in-time autotuning infrastructure is relevant for optimizing computation kernels that will be called numerous times with similar parameters through the execution, re-optimizes kernels when they are called with other parameters, and the programmer can obtain the optimal parameters to use them for other kernels. We present an experimental performance evaluation of our approach. Compiling the code introduces an overhead on the first iterations, and this overhead is compensated for during subsequent iterations. We also determined that the optimum found seems stable and accurate.
\end{abstract}

\maketitle
\keywords{LaTeX template, ACM CCS, ACM}

\section{Introduction}
\label{sec:intro}

Performance portability is a major concern on current architectures \cite{dubey2021performance}. Nowadays's compilers can provide very good support and, as presented in \cite{clangjit}, just-in-time (JIT) compilation can be a useful approach by providing the compiler information only known at run-time to allow for more accurate optimizations. 

Moreover, the architecture complexity makes tuning especially tricky. Performance can depend on parameters such as block size, loop unrolling factor, or even implementation choices such as loop ordering, fusion, or fission, to name a few. 

Pragmatic approaches to determine the best parameter set such as autotuning are attractive because they tune the program for a specific machine with a purely observation-oriented objective, in the sense that they make their decision based on the performance they observe. \emph{Offline} autotuning takes an isolated kernel, determines the optimal tuning parameters, and subsequent executions of the program use the optimized kernel. On the other hand, \emph{online} autotuning determines the optimal version at run-time.

Offline autotuning tunes a function once and for good: the optimal parameters found for the function can be used for any program that calls this function. On the other hand, online autotuning tunes a function for a given execution and a given program only: the parameters found cannot be reused for another program or even another execution of the same program. But online autotuning presents multiple advantages. Among them, we can mention the fact that it does not require any preliminary preparation, isolation, and setup of the kernels to be optimized. Moreover, the functions to optimize are optimized in the same conditions as the conditions of the execution: for instance, if multiple threads are competing for shared resources such as cache memory or the system bus, the optimal parameters might not be the same as those obtained during an execution on a machine dedicated to running this single computation kernel. 

A notable example of run-time tuning is the approach followed by FFTw \cite{frigo1998fftw}. Before calling the function that actually computes the FFT, the programmer is required to create a \emph{plan}, providing the size of the array on which the FFT will be computed, and the direction of the FFT. During this plan creation, the best possible execution options are determined, either by measuring or by estimating them. The latter approach is faster but expected to be less accurate. The former approach actually runs the various implementations of the FFT available for the requested plan size, measures them, and selects the fastest one. It is expected to be more accurate but takes some time. However, the FFT is expected to be called numerous times on data of the same size, so this plan creation time is assumed to be negligible compared to the performance gain obtained by this run-time best implementation selection.

FFTw's approach requires calling a function that performs not only the initialization of the plan but also this tuning. Our goal is to design a tool that requires as little intervention in the code as possible, in line with previous works that extended the C++ language for just-in-time compilation. As a consequence, we chose to work in the compiler and rely on a just-in-time compilation extension for C++. 

An alternative option would be to generate all the variants at compile-time, and only run and select the best one at run-time. This approach would avoid having to compile the code at run-time, but we would lose the advantages of just-in-time compilation. Our work combines both just-in-time compilation (and the fact that some parameters are known at run-time), and autotuning.

In this paper, we are presenting an infrastructure based on ClangJIT (which is based on LLVM) for just-in-time autotuning. It adds the optimization capabilities of an autotuner to JIT compilation and generates variants on-the-fly during the first calls to a function.

The remainder of this paper is organized as follows: Ssection \ref{sec:related} gives an overview of previous works on online and offline autotuning, and compares them to our approach. Section \ref{sec:design} describes the design of our autotuning infrastructure, and how we modified the compilation process to autotune functions at run-time. Section \ref{sec:perf} evaluates the performance of our approach. Last, \ref{sec:conclu} concludes the paper and mentions some open perspectives.

\section{Related works}
\label{sec:related}

Our work relies on ClangJIT and its just-in-time compilation capabilities \cite{clangjit}.  ClangJIT extends the C++ language with the {\tt [[clang::jit]]} attribute on function templates. Such function templates are then parsed, and an AST is generated at compile time, but compiled only at run-time. The template function can be specialized with type parameters, but also with non-type parameters that can be non-constant expressions. Henceforth, when this function is compiled at run-time, some parameters only known during the execution are known, and the optimization passes have more information. For instance, listing \ref{lst:related:clangjit} shows an implementation of the traditional {\tt saxpy} kernel, which computes $Y=a*X+Y$ for two vectors $X$ and $Y$ and a scalar $a$. In our code, the size of the vectors is passed as a non-type template parameter.

\begin{lstlisting}[frame=single, language=C++, basicstyle=\small, caption={Code example with ClangJIT}, label={lst:related:clangjit}]
template <typename T, int size>
[[clang::jit]] void mysaxpy( T a, T* x, T* y ){
  for( auto i = 0 ; i < size ; i++ ){
    y[i] = a * x[i] + y[i];
  }
}
\end{lstlisting}

Auto-tuning systems can be categorized into two categories: offline auto-tuning is made before the execution, for instance, to optimize a computation kernel that will be called with some known parameters on a given architecture; online auto-tuning is, on the other hand,  optimized at run-time. These systems are usually \emph{feedback-driven}, which means they follow a pragmatic approach that executes the program to be optimized and tries to improve it based on the information collected during the execution. It is worth noting that the objective of the optimization can be an execution time, but also something else, such as energy consumption, or even a combination of several ones for multi-objective optimization.

The Automatically Tuned Linear Algebra Software (ATLAS) project \cite{ATLAS} provides a broad set of linear algebra operations and optimizes them at install time. Linear algebra routines can be implemented using multiple algorithms and parameters. Which algorithm and parameter set are the best often depends on parameters such as the hardware it is executed on and the size of the matrices or vectors they are used on. ATLAS provides search scripts that try them and keep the fastest parameters for each tested configuration. Then at run-time, calls to the routines select the chosen implementation and parameters depending on parameters such as matrix sizes.

In \cite{kisuki2000combined}, the notion of \emph{iterative compilation} is explored: several versions of a program are generated, and their execution time is measured to keep the fastest variant. It is an example of pragmatic optimization. This work relies on a source-to-source compiler (MT1) that generates the variants. However, in this approach, the program is compiled at every step of the optimization process: it is an offline auto-tuning framework that executes the whole program. In \cite{pouchet2008iterative}, only the kernel code is generated and recompiled. This work focuses on optimizing loop nests in the polyhedral model, targetting programs that yield a large parameter space.

Active Harmony provides an API for automatic tuning \cite{activeharmony}. It can be used for both offline auto-tuning, for instance, with the loop optimization framework CHiLL \cite{tiwari2009scalable}, and online tuning \cite{tiwari2011online}.  The latter is closer to our work, which focuses on online tuning. The program to be optimized connects to a server that generates code variants corresponding to parameter sets. These code variants are compiled into a shared library, and the executed program loads them using {\tt dlopen} and {\tt dlsym}.

CHiLL is a code transformation framework focusing on loop optimization, working both at AST level and on the polyhedral model \cite{chill}, that can also generate parallel code for GPUs \cite{cudachill}. It can generate code variants as part of an auto-tuning workflow.

The Orio framework \cite{orio} implements an offline auto-tuning workflow that can be extended easily. The computation kernels to optimize are annotated with a dedicated language that defines the parameters and their possible values. Multiple search algorithms are available to explore the parameter space. For each point to be evaluated, a code variant is generated, potentially with a dedicated tool such as CHiLL \cite{oriochill}, executed and measured.

atJIT is probably the closest to our current work. It uses a Clang plugin (based on Easy::JIT \cite{easyjit}) and specialized compiler optimizations incrementally by trying to find optimal parameters such as loop optimizations (unrolling, tiling, etc), inlining... \cite{atjit}. It provides a set of parameter search algorithms, and the programmer has access to tuning parameters. It requires some modification in the source code, as shown by listing \ref{lst:related:atjit}: it uses a driver to maintain the state of the tuner; this driver has a method {\tt reoptimize} that returns either the optimal version or an optimized version of the function. The work we present in this paper is more integrated in the compiler to require fewer modification in the source code.

\begin{lstlisting}[frame=single, language=C++, basicstyle=\small, caption={Code example with atJIT}, label={lst:related:atjit}]
tuner::ATDriver AT;

const int M, N, K = ....;
for (...) {
  auto const& tuned_matmul = 
         AT.reoptimize(M, N, K, _1, _2, _3 );
  tuned_matmul(C,data(), A.data(), B.data());
  // ... A, B, C change
}
\end{lstlisting}


In addition to \emph{how} the application or the computation kernel is optimized, the size of the parameter space is a major challenge. While it is an important issue for offline auto-tuning, it is a major one with online auto-tuning, since the parameter search is included in the application's execution time: if the application needs to evaluate too many suboptimal points, or if the point search itself takes a significant time, the JIT tuning overhead will be higher, and might overcome the performance gain \cite{chen2015angel}. We think that all the works cited in this section provide interesting approaches, that are worth investigating in the context of our work, but outside of the focus of this paper.
\section{Design}
\label{sec:design}

\subsection{ClangJIT}
\label{sec:design:clangjit}

In \cite{clangjit}, Clang was extended to introduce the \verb|[[clang::jit]]| attribute on function templates. When a function is meant to be compiled at run-time, Clang keeps its (compressed) Abstract Syntax Tree (AST) without specializing the template. When this function is called in the code, a function \verb|__clang__jit| is called. It specializes the template (\emph{i.e.}, creates a version of the function using the template parameters), compiles it, and returns a function pointer. 

Non-type template parameters are then known at JIT-compilation-time, which allows for optimizations that would not be possible with ahead-of-time compilation: for instance, vectorization, loop unrolling, or some polyhedral optimizations can become possible.

\paragraph{Ahead-of-time compilation} At compile-time, Clang generates the abstract syntax tree (AST) of the function to be JIT-compiled. It is serialized and saved along with data such as the source file (useful to provide information to the developer).

\paragraph{Runtime libraries} ClangJIT is linked against a large runtime library that contains Clang and LLVM. It provides the function \verb|__clang_jit|, and keeps a cache of generated instantiations.

\paragraph{Just-in-time compilation} ClangJIT modified the code generation system so that when a JIT-compiled function is called or when a function pointer is retrieved, it actually calls \verb|__clang_jit|. This function provided by ClangJIT's runtime environment returns a pointer to the new function by looking up into the cache of instantiations. If this instantiation does not exist, the template function is specialized, compiled, and merged with externally-available definitions (and inserted into the cache).  

This compilation is protected by a mutex so that no concurrent compilations of the same function can be performed. 

\subsection{Just-in-time autotuning}
\label{sec:design:tuning}

Our contribution is built on ClangJIT, in which we modified the code generation workflow to insert autotuning capabilities. Our goal was to provide an accurate just-in-time tuning infrastructure, provided by the compiler, requiring only a few modifications in the source code and consistent with the template-oriented syntax followed by ClangJIT.

\paragraph{Overview}
We introduced a new keyword: \verb|__autotune__|. Template parameters passed with this keyword are arrays and contain the possible tuning parameters. We can see an example of how these tuning parameters can be used 

Listing \ref{lst:design:parameters} shows how a function can use these parameters, in this example with a blocking parameter. An array \verb|Y| is passed as an \verb|__autotune__| template parameter. The first times this function \verb|foo| is being called, it is specialized with the next element of this array and compiled. The execution time is measured, and once all the elements of the parameter array have been tried, the best specialization is kept.

\begin{lstlisting}[frame=single, language=C++, basicstyle=\small, caption={Using JIT tuning parameters}, label={lst:design:parameters}]
template <__autotune__ int** y>
[[clang::jit]] void foo( double* tab, int len, 
                         double alpha){    
    for(int i = 0;i < len; i+= *y){
        for(int j = i; j < min(len, *y+i) ; j++){
            tab[j] *= alpha;
        }
    }
}
\end{lstlisting}

Listing \ref{lst:design:passing} shows how this array of tuning parameters can be passed to the function call with the array \verb|z|.

\begin{lstlisting}[frame=single, language=C++, basicstyle=\small, caption={Passing JIT tuning parameters}, label={lst:design:passing}]
int z[3] = { 2, 4, 8 };
for(int i = 0 ; i < NB_ITER; i++) {
    foo<&z>();
}
\end{lstlisting}

\paragraph{Online autotuning} 
As we mentioned before, we are here following an \emph{online} autotuning approach, meaning that the generation of the variants and the selection of the best one are performed during the execution of the program. This generation and selection can be done during an initialization phase, similarly to what is done in FFTw. We chose to do it during the first calls to the function. The first time the function is called, it is generated and executed with the first autotuning parameter, and so on for each parameter. The goal is to avoid the initialization overhead caused by running the function just to measure it, and to optimize it on real data used by the program without the need for a deep copy of the data on which measurement runs will be made. Considering that we ahve $N$ varians and the function will be called $M$ times, instead of generating, compiling, and running the function $N$ times if we have $N$ possible variants and then executing $M$ times the optimized function, we are generating, compiling, and running a variant the first $N$ times, and then we are running the $M-N$ remaining calls using the optimized function.

\paragraph{Generating variants} 
If the autotuning parameter array contains $N$ elements, the first $N$ times the function is being called, it is instantiated with the next available parameter. Instead of inserting the function pointer in an LLVM DenseMap of generated functions, we are keeping the tuning information and the execution time of the best execution in another DenseMap. Once all the possible parameters have been tried, we are generating the best specialization one last time, and it is compiled and inserted info \verb|__clang_jit|'s cache of instantiation, as being done by ClangJIT. This final compilation is necessary because we can only keep ASTs and specialization data, and not the binary compiled by LLVM.

\paragraph{Autotuning parameters} 
ClangJIT allows integer non-type template parameters. Integer tuning parameters are relevant because they can be loop unrolling factors, block sizes... If required by applications, we can extend it to support other data types.

\paragraph{Choosing between implementations} 
Our autotuner can choose between several functions, as shown in Listing \ref{lst:design:function}. This can be useful, for instance, to choose between several implementations of a given operation. An array of the possible function pointers is passed as a parameter to a proxy template function that actually calls the function to run. This proxy function has the \verb|[clang::jit]| attribute and gets the array of possible indices as a template parameter. Following the same logic, every time the function is called, a specialization of this template is generated with the next index, therefore calling the corresponding function from the array of function pointers. When all the possible indices have been tried, this proxy function is generated and compiled and calls the best implementation.

\begin{lstlisting}[frame=single, language=C++, basicstyle=\small, caption={Passing JIT tuning parameters}, label={lst:design:function}, float=*]
template<__autotune__ int** id_param>
[[clang::jit]]launch_matmul(void (*funcArray[])(int**, int**, int**, int),
                                int ** A, int ** B, int **C,int N) {
    funcArray[*id_param](A, B, C, N);
}
int main(){
/* [...] */
    void (*matmulFunctions[3])(int**, int**, int**, int) = {matmul1, matmul2, matmul3};
    int tab[3] = {0,1,2};
    launch_matmul<&tab>(matmulFunctions, A, B, C, N);
/* [...] */
}
\end{lstlisting}

\paragraph{Handling calls with different arguments}
The optimization of a function is very data-dependant: the best optimization parameters found for, for instance, a given data size are likely to be different from the best ones for another data size. We give the programmer two options regarding optimization. 

The state of the autotuning process is kept for a given function with a given optimization variable: we are keeping the name of the autotuning template parameter in the map that keeps the tuning parameters and the performacne obtained with them. If this parameter's name changes, we consider it to be another autotuning problem and another instance of the autotuner is being created to start the autotuning process from 0.

However, if the programmer considers that the optimal parameters found for a function are valid for other function calls, they can keep them and use them for other function calls. For instance, Listing \ref{lst:design:parameter} presents a basic matrix-matrix multiplication with loop tiling. This function can return the value of the autotuning parameter that has been used. After the tuning iterations, only the optimal value found by the autotuner will be used, so the calling program can use it for other JIT-compiled functions. For instance, if the programmer decides that this block size can be used by other computation routines, they can define these routines as JIT-compiled templates and pass it as a non-type template parameter.

\begin{lstlisting}[frame=single, language=C++, basicstyle=\small, caption={Retrieving JIT tuning parameters}, label={lst:design:parameter}, float=*]
template <__autotune__ int** block_size>
[[clang::jit]]int multiplyMatrixByBlocks(int** A, int** B, int** C, int n) {
    for (int i = 0; i < n; i += *block_size) {
        for (int k = 0; k < n; k += *block_size) {
            for (int j = 0; j < n; j += *block_size) {
                for (int ii = i; ii < min(i + *block_size, n); ++ii) {
                    for (int kk = k; kk < min(k + *block_size, n); ++kk) {
                        for (int jj = j; jj < min(j + *block_size, n); ++jj) {
                            C[ii][jj] += A[ii][kk] * B[kk][jj];
                        }
                    }
                }
            }
        }
    }
    return *block_size;
}
\end{lstlisting}



\paragraph{Performance measurement} 
We chose to measure the execution time of the generated function by counting CPU cycles with \verb|rdtsc|. However, this function can be overloaded and any other measurement function can be used to count any other metric, such as energy consumption.

\subsection{Compilation overhead versus performance gain}
\label{sec:design:overhead}

In our approach, the first iterations require compiling and executing every variant of the function. These variants contain the fastest and the slowest implementation available. 

If we consider an equal cost for each compilation, denoted $C$, and that the execution time of the $k$ variants is denoted ${E_0, E_1,..., E_{k-1}}$ ($E_0$ being the fastest and $E_{k-1}$ being the slowest), the total execution time of $N$ iterations will be:

\begin{align}
    E_{auto} &= \sum_{i=0}^{k-1} (C+E_i) + C+E_0 + \sum_{i=0}^{N-k-1} E_0\nonumber \\
    &= kC+\sum_{i=0}^{k-1} E_i + C+(N-k-1) E_0\label{eq:execauto}
\end{align}
 
Let's assume that without autotuning, the programmer would have picked up an implementation $p$ which has an execution time denoted $E_p$. Therefore, we want to be in conditions where $E_{auto}$, given by equation, \ref{eq:execauto} is bounded by $NE_p$:

\begin{align}
N\cdot E_p \geq \sum_{i=0}^{k-1} (C+E_i) + C+E_0 + \sum_{i=0}^{N-k-1} E_0 \nonumber \\
\Leftrightarrow (N-k) (E_p-E_0) \geq (k+1)C + \sum_{i=0}^{k-1} E_i - kE_p \label{eq:execauto2}
\end{align}

The left part of equation \ref{eq:execauto2} is the performance gain obtained on the last $N-k$ iterations (after the tuning has been done). The right part is the overhead caused by the compilation of the first $k+1$ executions and the difference between $E_p$ and the execution time of the other variants ($\sum_{i=0}^{k-1} E_i)$ can be larger or smaller than $kE_p$). 

So in other words, we need to (a) optimize quickly (to have $k << N$), (b) have only a few slow variants (to avoid having $\sum_{i=0}^{k-1} E_i)\ >> kE_p$), and (c) get a significant performance gain (to have a large $E_p-E_0$).



\section{Performance evaluation}
\label{sec:perf}

We implemented our work in ClangJIT and we compiled it with GCC 11.4. Our experiments ran on a machine featuring 2 AMD EPYC 7763 processors, for a total of 256 threads, and 1 TB of RAM. We compiled LLVM in Release mode, and we compiled the application programs with -O3.

\subsection{Consistency of choice}
\label{sec:perf:blocksize}


We ran the computation kernel presented on Listing \ref{lst:design:parameter}, modified to return the block size used by the function. As we mentioned in section \ref{sec:design:tuning}, on the first iterations (the \emph{tuning} iterations), the current parameter is being used, then the one that gives the fastest result is kept for the remaining iterations. 

\begin{figure}
    \centering
    \includegraphics[width=\linewidth]{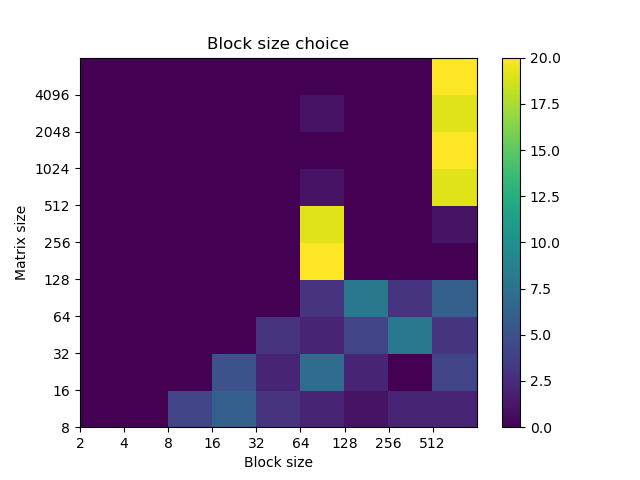}
    \caption{Block size choice for different matrix sizes.}
    \label{fig:perf:blocksize}
\end{figure}

Figure \ref{fig:perf:blocksize} shows the number of times each block size was chosen. We can see that 64 was always chosen for medium-size matrices (128, 256), while 512 was always chosen for larger matrices (512 and larger). For smaller matrices, the benefit of loop tiling is not obvious and the results vary because the execution times are close to each other regardless of the block size.

This result is conditioned by the reproducibility of the performance measurements. If no execution of the function stands clearly as the best one, the chosen parameter will vary between executions of the program. But in this case, any of the best parameters can be used. With our benchmark, we can see that some block sizes are distinctly better than others and that this block size depends on the size of the matrix: the function cannot be optimized once, but the optimization needs to be performed for different matrix sizes.

\subsection{Overhead on the execution}
\label{sec:perf:overhead}


During the first executions, the optimized function is generated and compiled at run-time, which introduces an overhead compared to regular, ahead-of-time compiled functions. This overhead is expected to be compensated by the performance gain on subsequent executions of this function. 

We executed a code that chooses between implementations of a straightforward matrix-matrix multiplication with three different loop orders (ijk, ikj, jik) as shown by Listing \ref{lst:design:function}, on 100 iterations of a loop that calls this matrix-matrix multiplication function and measures its execution time. The function was JIT-compiled three times: on the first three iterations (the \emph{tuning} iterations), and on the fourth one which compiles the final version. 

\begin{figure}
    \centering
    \includegraphics[width=\linewidth]{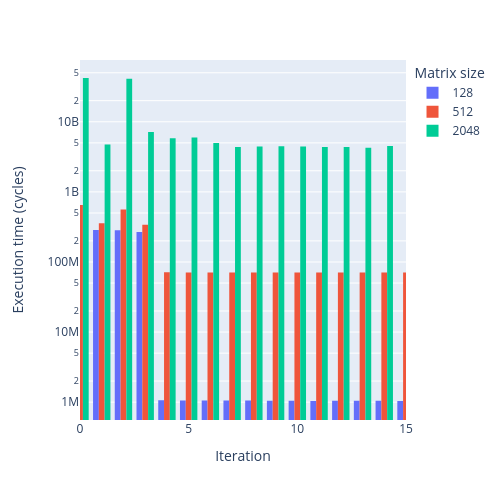}
    \caption{Execution time for each iteration, for three matrix sizes (logarithmic scale on the Y axis).}
    \label{fig:perf:iteration}
\end{figure}

Figure \ref{fig:perf:iteration} shows the iteration time for each iteration (on the first 15 iterations), using three matrix sizes. As expected, the compilation time introduces a larger relative overhead with small matrices (blue bars), and it is small on larger matrices. Iterations 0 and 2 used slower implementations, so they take significantly longer than the other iterations. Iteration 1 used the fastest implementation, and we can see that, on larger matrices, the overhead on this iteration is very small. 

\subsection{Overhead amortization}
\label{sec:perf:amort}

The general principle of our approach is that we accept the cost of compiling the function at run-time because this overhead is expected to be compensated for by the performance gain obtained during the other calls to this function by the fact that the function is optimized. 

In section \ref{sec:design:overhead}, we examined the conditions for the autotuning process to yield performance gains. We concluded that the performance gain must be significant enough to surpass the compilation cost.

We executed the benchmark that chooses between matrix-matrix multiplication functions on 100 iterations, and we measured the cumulative execution time for several matrix sizes. We executed the benchmarks 300 times on small matrices to mitigate potential variability due to system disturbance, and 12 times on large matrices. 

\begin{figure}
    \centering
    \includegraphics[width=\linewidth]{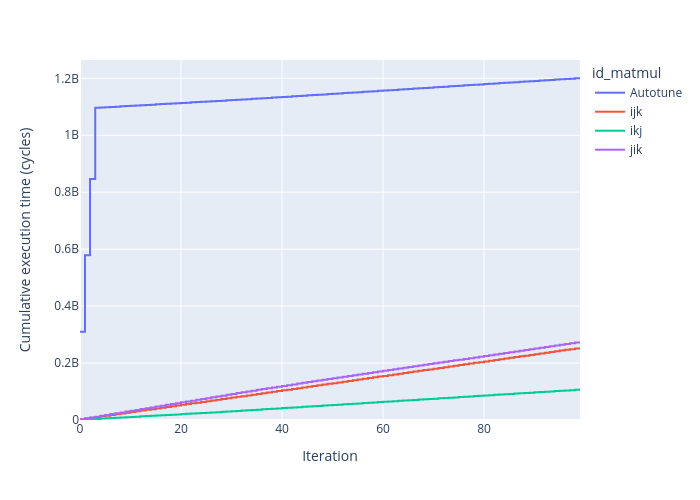}
    \caption{Cumulative execution time, small matrices (N=128).}
    \label{fig:perf:cumul128}
\end{figure}

Figure \ref{fig:perf:cumul128} shows the execution time of the autotuned version and the three algorithms on matrices of size $128\times128$. We can see that although the slope of the autotuned curve is the same as the best one, the compilation cost is prohibitive and surpasses the performance gain on these 100 iterations. For the autotuned curve to cross the non-optimal ones, the program would need to call this function a larger number of times. As a consequence, we observe that with this small matrix for which the performance gain is not significant enough, the compilation cost is high compared to the execution time, and our approach would be valid only for programs that call the function a very large number of times.

With larger matrices, the compilation cost is relatively small compared to the execution time and the overhead caused by the tuning iterations is limited. We can observe that the autotuned curve has the same slope as the optimal one, shifted up by a time corresponding to this tuning overhead.

The performance gain on larger matrices is significant enough for the performance curves to diverge quickly, and the translation due to the optimization overhead is small compared to the execution time. As a consequence, our autotuned version gives better performance after only a few iterations, and the performance loss compared to a very skilled programmer that would optimize the code directly is small.

\begin{figure}
    \centering
    \includegraphics[width=\linewidth]{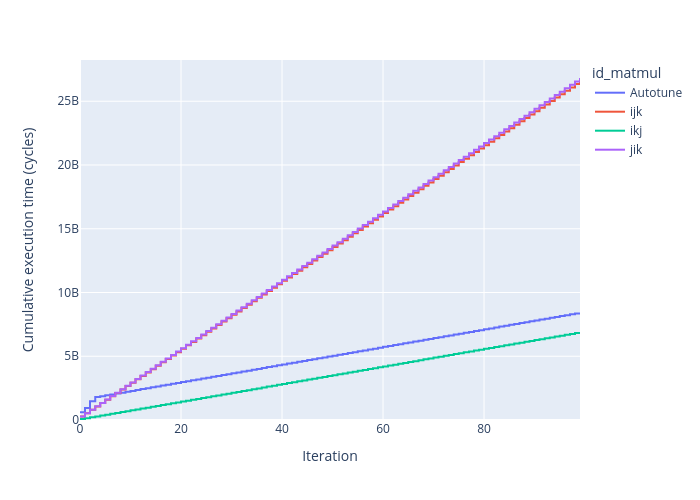}
    \caption{Cumulative execution time, medium-size matrices (N=512).}
    \label{fig:perf:cumul512}
\end{figure}

\begin{figure}
    \centering
    \includegraphics[width=\linewidth]{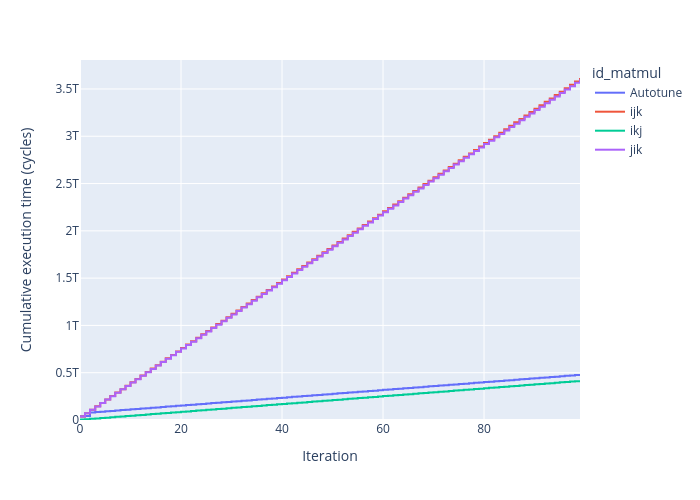}
    \caption{Cumulative execution time, larger matrices (N=2048).}
    \label{fig:perf:cumul2048}
\end{figure}

\section{Conclusion and perspectives}
\label{sec:conclu}

In this paper, we have presented a just-in-time (JIT) autotuning approach integrated into a JIT-compilation extension of C++.  We show that compiler support allows for run-time autotuning with little modification in the code. An experimental evaluation using benchmarks shows that this approach gives good results easily, as long as the potential performance gain on the subsequent iterations is not too small.

It must also be noted that we do not modify the program's behavior, in the sense that we do not produce any code modification: we are selecting the best option between those provided by the programmer.

Since this approach gives promising results, it opens perspectives in the field of optimization. First, we only looked at code parameters (block size, loop unrolling, choice of function...). Since we are working in the compiler, we can extend our infrastructure to support compiler options for optimization.

With a larger parameter space, parameter sweep is not a viable option because it requires a large number of tuning iterations. Heuristics for faster convergence to the optimum exist in the literature, such as those mentioned in section \ref{sec:related} or works such as \cite{menon2020auto}\cite{wu2022autotuning}, and it would be interesting to implement and evaluate them in our infrastructure.

Finally, we are going to put together a portfolio of applications and autotune them using our approach. Our goal is performance portability, and we expect our approach to spare us from making invasive modifications to the application code. We can cite \cite{wu2021performance} that presents optimization strategies for the ECP proxy app SW4lite or \cite{karlin2013tuning} that presents the optimizations that were made on LULESH to adapt it to new hardware. It would be interesting to see if our JIT autotuner can get the same results with as few modifications to the code as possible.

Our implementation is available on GitHub: \url{https://github.com/MorelElian/JIT_autotuning}

\section{Acknowledgments}

We would like to thank Hal Finkel for the discussions we had on ClangJIT, and Calcul Québec for their support in this project.

\bibliographystyle{ACM-Reference-Format}
\bibliography{refs}

\end{document}